\documentclass[showpacs,showkeys,floatfix,prb,twocolumn]{revtex4}

\usepackage{graphicx}
\usepackage{dcolumn}
\usepackage{bm}
\usepackage{setspace}
\usepackage[normalem]{ulem}
\newcommand{\bea}{\begin{eqnarray}}
\newcommand{\eea}{\end{eqnarray}}
\newcommand{\be}{\begin{equation}}
\newcommand{\ee}{\end{equation}}
\newcommand{\la}{\langle}
\newcommand{\ra}{\rangle}
\newcommand{\half}{\frac 1 2}
\begin{document} 

\title{Elementary Excitations in Bose-Einstein Condensates at Large
  Scattering Lengths}
%
% Text modified by Sarjonen, original obtained from Saarela on Monday 16th.
%
\author{R.\ Sarjonen and M.\ Saarela}
\affiliation{Department of Physics, P.O. Box 3000, FIN-90014, 
University of Oulu, Finland}

\author{F. Mazzanti}
\affiliation{Dep. de Fisica i Enginyeria Nuclear,                             
Universitat Politecnica de Catalunya,                            
Campus Nord B4-B5, E-08034 Barcelona, Spain
}

\begin{abstract}

  We present a theoretical analysis of excitation modes in
  Bose-Einstein condensates in ultracold alkali-metal gases for large
  scattering lengths 
  and momenta
  where corrections to the mean field
  approximation become important.  We assume that the effective
  interaction in the metastable, single channel, gaseous phase has a
  well defined Fourier transform that scales with the scattering
  length. Based on this we show that for increasing scattering lengths
  or equivalently increasing densities the system becomes less
  correlated and that at large values of 
  the scattering length Bragg
  scattering measures directly the Fourier transform of the effective
  two-body potential. We construct model potentials which fit the
  recently measured line shifts in $^{85}$Rb by Papp et
  al. (Phys. Rev. Lett. {\bf 101}, 135301 (2008)), and show that they
  fix the low momentum expansion of the effective range function. 
  % Here i soften a tad the statement about us finding an r_eff<<1 is
  % the way to go. I have changed it a bit in order to make it sound less
  % as if this was a new result from us, while it was actually derived
  % by Bruun et al.  
  We find excellent agreement with the experimental data when 
  the effective range is
  $\ll 1$ and the coefficient of the
  $k^4$-term is $-7.5 \pm 0.5$ in scattering length units.  The
  resolution in Bragg scattering experiments so far does not reveal
  details of the frequency dependence in the dynamic structure
  function $S(k,\omega)$ and we show that the Feynman spectrum
  determines the measured line shifts. We propose the 
  possibility of a
  transition to a novel density wave state.
\end{abstract}

\pacs{02.70.Ss} \keywords{Bose-Einstein condensate}

\maketitle

% Introduction to experiments and recent theories

Bragg spectroscopy measurements of excitation spectra in ultracold
atomic gases with Bose-Einstein condensation have 
shown excellent agreement with the predictions of the
 Bogoliubov theory of elementary
 excitations~\cite{StamperKum1999,Steinhauer2002,Ozeri2005,NatureErnst2009}. 
However, in recent experiments \cite{Papp2008} on $^{85}$Rb particle
density, scattering length and transferred momentum have reached such high values
that deviations from that are clearly observed and a possible relation
to the  $^4$He-like roton behavior has been speculated.

From the theoretical side the dynamic structure function $S(k,\omega)$
determines the excitation modes of a quantum Bose system. At low momenta
and frequencies typical Bose systems show
a sharp peak which defines the elementary
excitation mode. 
%In helium that mode has the familiar phonon-maxon-roton structure. 
At given k but higher frequency,
multiphonon excitations form a broad continuous distribution. In the line shift experiments \cite{Papp2008},
the resolution is not high enough to separate out these
contributions. Consequently, the experiments reveal information only
on the average structure. For a fixed k,
$S(k,\omega)$ satisfies strict sum rules 
\begin{eqnarray}
m_0(q) \!\! & = & \!\!\! \hbar\int_0^\infty d\omega\,S(k,\omega) = S(k) \ ,
\nonumber \\ 
m_1(a) \!\! & = & \!\!\! \hbar^2\int_0^\infty d\omega\,\omega\,S(k,\omega) = 
{\hbar^2 k^2 \over 2m } \ ,
\end{eqnarray}
with $S(k)$ the static structure factor.
These sum rules implies that the average
value of the frequency $\la \omega \ra=m_1(q)/(\hbar m_0(q))=\hbar k^2/(2 m S(k))$
obeys the Feynman spectrum.
%\be
%\la \omega(k)\ra =\frac{}.
%\label{Feynman-ek}
%\ee
In the analysis of experiments\cite{Papp2008} the frequency
dependence of $S(k,\omega)$ was fitted with a Gaussian
function. 
% I think this defined the way it is done IN THE EXPERIMENTS. 
% In theretical calculation I would more think about finding a sharp
% peak in S(k,w). But experiments do kill all detail in S(k,w) so that
% is it.
There, line shift at a given momentum $k$ is obtained
as the deviation of
the maximum of the broad frequency distribution of the measured
$S(k,\omega)$ from the free particle value $\hbar^2k^2/(2m)$. If we
identify the maximum with the average value $\la \omega(k)\ra$,
the line shift can be calculated from purely ground state quantities
\be
\Delta \omega(k) = \frac{\hbar k^2}{2 m}\left(\frac 1 {S(k)}-1\right)\,.
\ee

% Introduction to our work
In this Letter we present results on the static and 
dynamic  structure functions of bosonic fluids using the
correlated basis function method\cite{FeenbergBook} widely applied to
strongly correlated fluids like the charged Bose gas\cite{bosegas} and in
particular superfluid $^4$He\cite{EKthree,Chuckphonon}. We assume that the
effective interaction scales with the scattering length $a$ by setting
the length unit equal to $a$ and the energy unit equal to $\hbar^2/(2
m a^2)$. Since we compare our results with measurements on $^{85}$Rb,
we set $m$ equal to the $^{85}$Rb mass. Experiments are done for 
the fixed momentum transfer $k=4\pi/780\,$nm and
density $\rho=7.6\times 10^{13}$cm$^{-3}$, but changing the
scattering length $a$ on a wide range  $153 a_B <a
< 891 a_B$ where $a_B$ is the Bohr radius. 
The only quantity
then needed to control 
the many-body problem is the {\it gas parameter} $x= \rho
a^3$.

Two simple models, which scale with the scattering length, have been
studied extensively in the
literature. The first one is the Fermi
pseudopotential that acting on non--singular wave functions has the
form $(4\pi a \hbar^2/m)\delta(r)$ with the strength proportional to
the scattering length. The second one is the gas of hard spheres where
the diameter of the particles is equal to the scattering length.
These models give the linear behavior for the line shift as a function
of the scattering length when $k a \ll 1$, but begin to deviate from each
other and also from the experiments when $k a> 0.4$.  That is why we
extend in this work the analysis to more realistic effective
potentials, which have finite range and finite strength, but assume
that they have a Fourier transform.
%For positive
%scattering lengths the effective potential can either be repulsive or 
%have bound states. However, bound states in the two-body interaction make
%Bose gases unstable against clustering, 
%liquid or solid formations. 
%The two-particle $^{85}$Rb-$^{85}$Rb
%potential in free space has many bound states and the many-particle
%ground state at $T=0 K$ is a solid. Here we want to analyse the
%effective interaction in the homogeneous, metastable, gaseous phase,
%where the Bose condensation is observed, assuming that the many-body
%wave function has no nodes.

% Here I add a paragraph and a figure discussing why we can d with
% teh Feynman approximation and don't really need to resort on the CBF
 \begin{figure}[t]
    \includegraphics[width=0.4\textwidth]{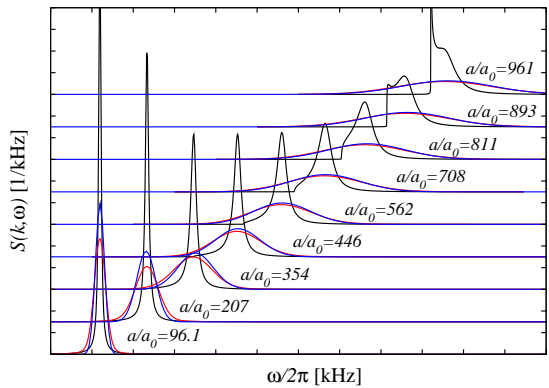}
    \caption{Comparison of the CBF $S(k,\omega)$ (black line), the CBF 
      $S(k,\omega)$ folded with the instrumental resolution functions of Ref.~\protect\onlinecite{Papp2008}
      (red line), and the Feynman approximation also folded with these
      functions (blue line) for soft spheres potential with the radius $R=3.5a$. Notice that
      the curves have been artificially shifted for clarity. }      
    \label{fig:CBFIR_figure}
  \end{figure}

The analysis of the line shift experiments requires a theoretical
model for $S(k,\omega)$ that allows finding the spectrum of
elementary excitations. A suitable choice for that is provided by
the Correlated Basis Function (CBF) method~\cite{Chuckphonon}, which
is very well suited to describe 
weakly interacting Bose gases\cite{bosegas}
and yields reasonable results also for $^4$He. We have found,
however, that due to poor instrumental resolution, the detailed
structure of $S(k,\omega)$ is washed out in the experiments and that
the elementary excitation spectrum is very well described by the
Feynman approximation. This is seen in Fig.~\ref{fig:CBFIR_figure},
which shows a CBF calculation for a system of bosons interacting
through the soft spheres potential $V(r\leq R)=V_0, V(r> R)=0$ for
$R=3.5$ in scattering length units.
We have found that this simple model describes fairly well
the experimental line shifts\cite{Papp2008}. The figure also shows
the CBF $S(k,\omega)$ and the Feynman approximation, both folded
with the instrumental resolution function of the experiments.  The
maxima of these broadened distributions agree very well and we
conclude that the relevant quantity describing the dynamics in the
experiment is $S(k)$.
 
We model the gas of $N$ $^{85}$Rb atoms
of mass $m$ with the Hamiltonian
\begin{equation}
H = -{\hbar^2 \over 2m} \sum_{j=1}^N \nabla^2_j +
\sum_{1=i<j}^N V(r_{ij}) \mbox{{\bf{,}}}
\label{hamil-a1}
\end{equation}
and the ground state   many-body wave function 
\begin{equation}
\Psi_0({\bf r}_1, {\bf r}_2, \ldots, {\bf r}_N) =
\prod_{1=i<j}^N f(\vert{\bf r}_i-{\bf r}_j\vert)
\label{jastrow-a1}
\end{equation}
is defined as a product of two--body correlation
functions\cite{FeenbergBook}, The unknown correlation function $f(r)$
is the solution of the variational problem which minimizes the
expectation value of the Hamiltonian.
%\begin{equation}
%\frac{ \delta \la\Psi_0|H|\Psi_0\ra}{\delta f(r)} = 0 \ .
%\label{eulerla-a1}
%\end{equation}
With diagrammatic techniques the resulting Euler-Lagrange equation can be written in a
Bogoliubov--like form for the structure function $S(k)$~\cite{EKthree}, which in scattering length units becomes
\be
S(k)=\frac{k}{\sqrt{k^2+2\tilde V_{\rm p-h}(k)}}\,.
\label{Sk}
\ee
The effective, many-body potential in coordinate space splits into two parts,
\begin{eqnarray}
\label{Vphr}
V_{p-h}(r)&=&V(r)+
\\ \nonumber
&+&\left[g(r)-1\right]\left[w_{\rm ind}(r) + V(r)\right] 
+2\left(\nabla\sqrt{g(r)}\right)^2\,. 
\end{eqnarray}
The potential on the first line is just the bare two-body
potential. In momentum space 
\be \tilde V(k) = \frac{4\pi^2 x}{k} \int
r dr V(r)\sin(k r)
\label{potFt}
\ee 
defines the 
{\it uniform limit approximation} (with $\tilde V_{p-h}(k)=\tilde V(k)$)\cite{FeenbergBook,Smith1979} for $S(k)$, 
that reduces in turn to the ordinary Bogoliubov expression when $V(r)$ is
further approximated by the pseudopotential. The rest of the terms in Eq. (\ref{Vphr}) contain the
many-body contributions to the effective interaction.
The radial distribution function $g(r)$ is
calculated from the inverse Fourier transform \be g(r) - 1 = \frac
{1}{2\pi^2x r} \int k dk (S(k)-1) \sin(k r).
\label{gFt}
\ee 
and the induced potential $w_{\rm ind}(r)$ is obtained by taking the
inverse Fourier transform of the expression
\be 
\tilde w_{\rm ind}(k)=-\frac{1}{2}\left(2 S(k)+1\right)\left(1-\frac{1}{S(k)}\right)^2\,.
\label{windk}
\ee
Eqs.\ (\ref{Sk})-(\ref{windk}) form a set of self-consistent equations
that have to be solved iteratively.

\begin{figure}[tbh]
\includegraphics[width=0.4\textwidth]{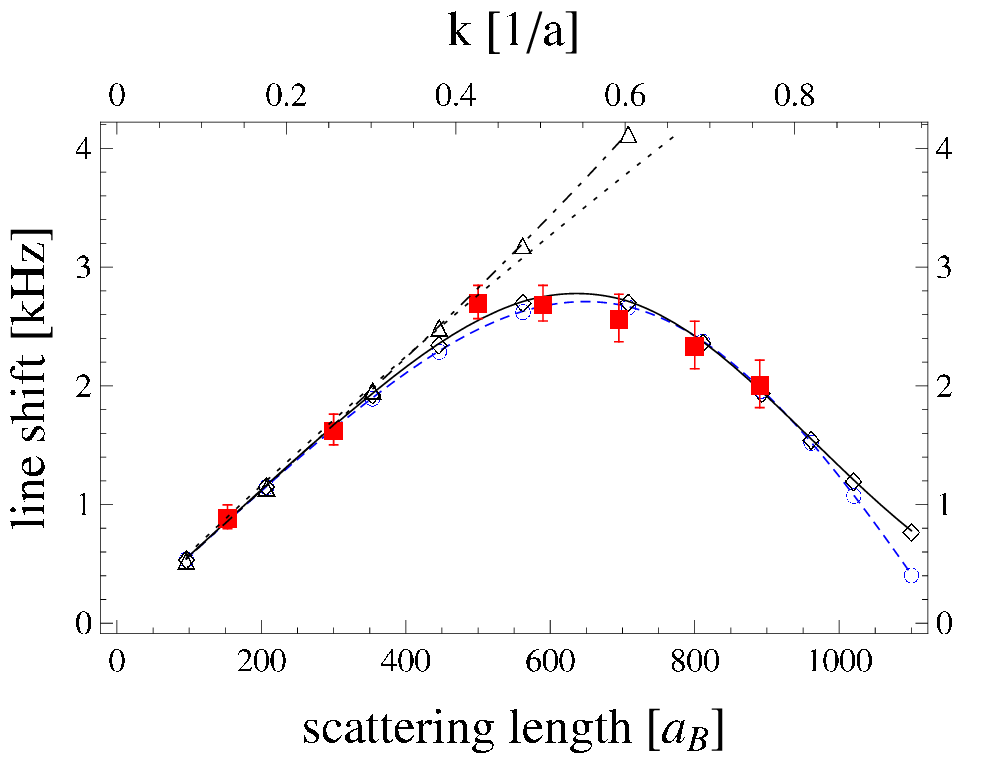}
\caption{The line shift $\Delta \omega(k)$ as a function of the
  scattering length calculated using the PTGT (solid line)
  and SW (dashed line) potentials. Theoretical results are compared
  with the experimental points with error bars of
  Ref. \protect\onlinecite{Papp2008}. Also shown are the hard sphere
  (dashed-dotted curve) and the pseudopotential (dotted
  curve) results. 
  The upper axis shows $k a$ for a momentum $4\pi /780\,$nm$^{-1}$.
% The momentum values given on the top axis 
%   are $4\pi a/780$.
%  \n{$4\pi a/780 ~{\rm nm}$.}
  }
\label{fig:Bragg_experiment}
\end{figure}

% %%%%%%%%%%%%%%%%%%%%%%%%%%%%%%%%%%%%%%%%%%%%%%%%%%%%%%%%%%%%%%
\begin{figure}[tbh]
{~~\includegraphics[width=0.38\textwidth]{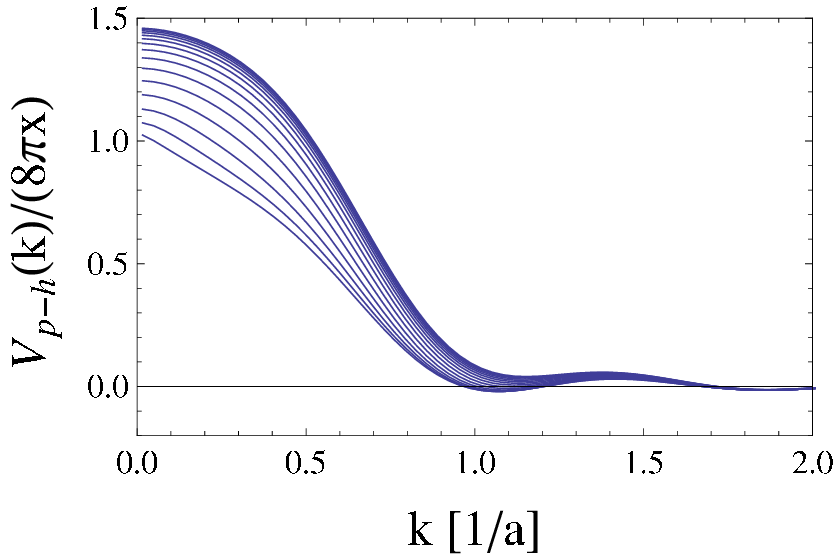}
\newline\newline
\includegraphics[width=0.38\textwidth]{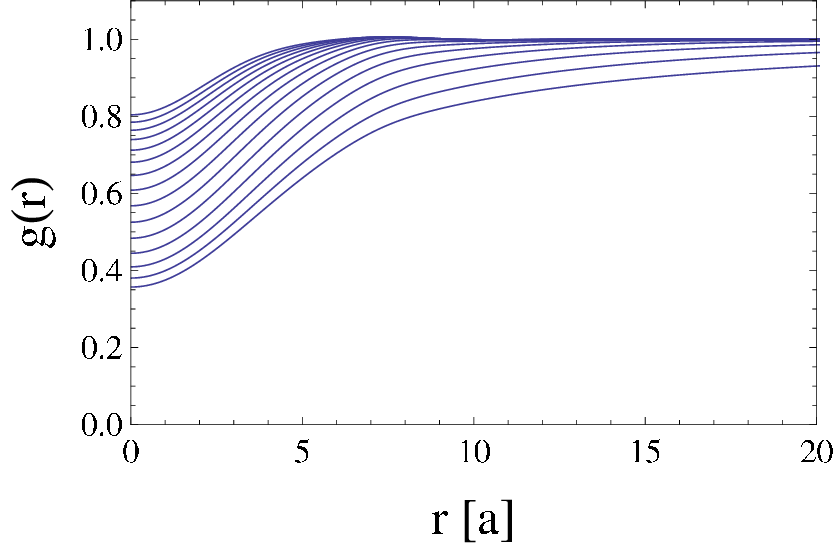}}
\caption{In the upper panel the effective many-body potentials as a
  function of momentum are shown for   the PTGT-potential 
  and different values of the scattering 
  ength   ranging from 100$a_B$ to 1500$a_B$ with
  100$a_B$ steps. The lowest curve is for the lowest value of $a$ and
  the highest curve is equal to the Fourier transform of the bare
  two-body potential $\tilde V(k)$.  
  The lower panel shows
  the radial distribution functions calculated using the
  PTGT-potential for the same values of the
  scattering length $a$.  The lowest curve is
  for the lowest value of $a$.  }
\label{fig:Vphk_figure}
\end{figure}
%% %%%%%%%%%%%%%%%%%%%%%%%%%%%%%%%%%%%%%%%%%%%%%%%%%%%%%%%%%%%%%% 

A measure of the strength of the
correlations in a quantum system is given by the amount $g(r)$
deviates from unity. From Eq. (\ref{Vphr}) we see that many-body
effects diminish when $g(r)\rightarrow 1$. 
That happens when $x$
increases for any potential with a Fourier transform as shown by
Eqs. (\ref{gFt}) and (\ref{potFt}) provided that $S(k)$ has no
singularities.
For the PTGT-potential we find $V_{p-h}(k)\approx V(k)$ for
$x\gtrsim 0.02$.
% $x\geq 0.02$.
% that happens when $x\geq 0.02$.

In the long wave length limit
\be
\tilde V_{\rm p-h}(k=0) =  m c^2
\left( \frac{2m a^2}{\hbar^2} \right)
\ ,
\label{sound} 
\ee
where $c$ is the speed of sound. At small $x$, i.e.\ $x<10^{-3}$, one
recovers the pseudopotential result $\tilde V_{\rm p-h}(0)=8\pi
x$. The agreement with Monte Carlo simulations \cite{giorgini-1} is
very good up to $x\approx 0.1$ even for the hard
spheres\cite{mazzanti-1}. For soft-core type potentials, the agreement
should be even better at large $x$, because many-body 
contributions to $V_{p-h}(k)$ become less important.

We use two models for the single channel potential. The first
one is the Poschl-Teller potential to which we have added
a weak, attractive, Gaussian tail (PTGT). The attractive tail mimics
the long range part of the van der Waals interaction,
and it improves the fit to the experimental line shifts,
\be
V(r) = \frac{G_1}{\cosh^2(\mu r)}-G_2e^{-\alpha(r-r_0)^2}\,,
\label{PTGT}
\ee 
where $G_1=0.306 038$, $G_2=0.012 955$, $r_0=7.377 861$, $\mu=0.383
580$, and $\alpha=4.799\mu^2$.  The second model in our
analysis is the step-well potential (SW)
\begin{equation}
V(r) = \left\{
\begin{array}{ll}
V_1, & \,\,\,\,\,\,r<R_1 \\
-V_2 & \,\,\,\,\,R_1<r< R_2 \\
0&\,\,\,\,\,r> R_2 
\end{array}
\right.
\label{SW}
\end{equation}
with $V_1= 0.110 786$, $V_2 = 0.010 572$, $ R_1 = 4.684 541$ and $R_2
= 1.5R_1$. All parameters are in scattering length units. They are fixed by setting the effective range $r_{\rm eff}/a \ll 1$, based on the multichannel
scattering analysis near the Feshbach resonance by Bruun et
al. \cite{BruunJackson2005} and by fitting the line
shift data.

In Fig. \ref{fig:Bragg_experiment} we compare our results to
experiments. Up to $a=400 a_0$ the behavior is almost linear and also
both the pseudopotential and the hard sphere models fit well with the
experimental data.  Those models, though, are unable to reproduce the
strong downwards bending observed in the line shifts at higher values
of the scattering length.
%That is why 
%the more flexible potentials of
%Eqs. (\ref{PTGT}) or (\ref{SW}) are needed.
%
\begin{figure}[tb]
\includegraphics[width=0.4\textwidth]{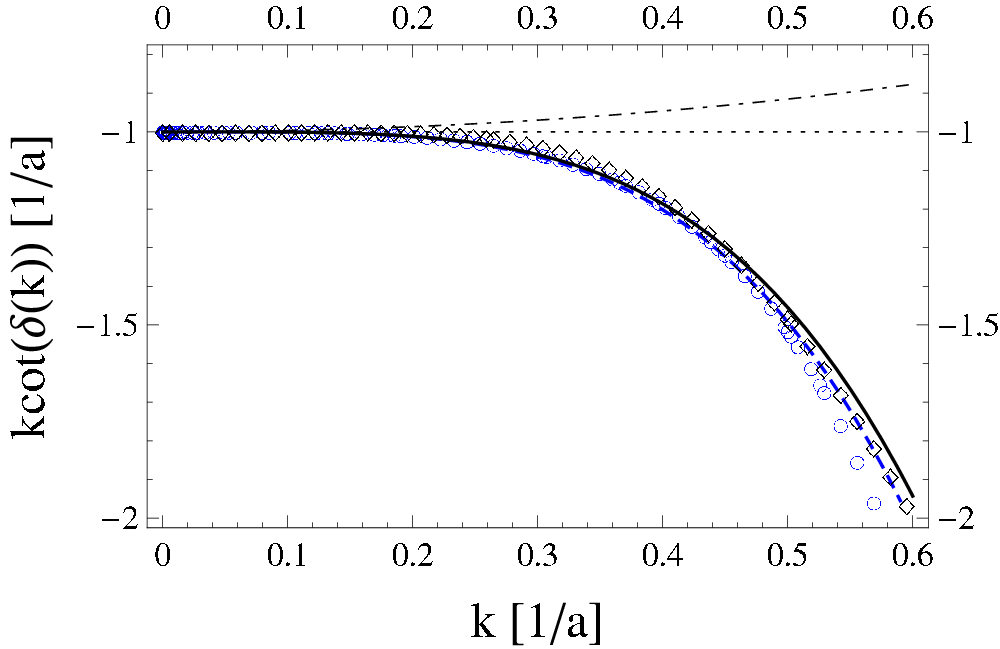}
\caption{The effective range function as a function of the two-body
  scattering momentum for the
  PTGT (diamonds) and SW (circles)
  potentials. Least square fits are done for the range $0<k<0.5$ to
  the PTGT (solid curve) and SW (dashed curve) potentials. Also shown
  are the results for the hard sphere (dashed-dotted curve) and
  $\delta$-function (dotted curve) potentials.  }
\label{fig:effrangefunc}
\end{figure}

The effective interaction $\tilde V_{\rm p-h}(k)$ determines the line
shifts completely within the resolution of the experiments. In
Fig. \ref{fig:Vphk_figure}a we show its behavior for different values
of the scattering length ranging from 100$a_B$ to 1500$a_B$ with 100$a_B$
steps using the PTGT-potential of Eq. (\ref{PTGT}). 
Since we have fixed the density to the experimental value,
that range corresponds to $10^{-5}<x<0.04$ in the gas parameter. In the figure we have divided
out the Bogoliubov constant $8 \pi x$ and that is why at $k=0$ the
lowest curve is very close to unity. 
Increasing the
scattering length increases the strength of the 
% Once again this we know excatly for our models...
potential at $k=0$
since the speed of sound increases (see Eq. (\ref{sound})). The upper
curves at the largest values of $a$ converge nicely to 
the Fourier transform of the bare potential. This means that
many-body effects in $V_{p-h}(k)$
vanish and therefore
the largest values available in experiments directly test the Fourier
transform of the effective, bare potential. 
As long as the system remains in the gaseous phase, the particle--hole
potential is directly given by the bare potential even if one keeps
increasing $a$.

The momentum transfer fixed in the experiments varies in $1/a$-units from
0.13 to 0.76 as depicted in
Fig. \ref{fig:Bragg_experiment}. Clearly understanding the behavior of
the effective interaction as a function of the
momentum transfer is essential since
at $k a \approx 1$ the effective potential 
approaches zero defining 
an inverse healing distance $\xi^{-1}$ 
where the line shift should also go to zero. 

The potential models chosen in Eqs. (\ref{PTGT}) and (\ref{SW})
give very similar line shifts within the experimental range, but at
larger $a$ they behave very differently. Oscillations evident in the
Fourier transform of the SW potential make it go through zero at $k a \approx 0.95$ 
whereas the PTGT-potential remains positive up to $ka\le1.7$. 
That difference could be experimentally tested by increasing the used momentum transfer by
$\sim$20\%. The SW potential also leads to the density wave
instability at $a\approx 2260a_B$ with $k a= 1.15$ whereas with the
PTGT-potential the system remains in the gaseous phase. It 
could also be interesting to search for that phase transition experimentally.

In order to 
better understand the importance of the correlations we have
plotted in Fig. \ref{fig:Vphk_figure}b
the radial distribution function $g(r)$ 
of the PTGT-potential for scattering lengths 
spanning the range $100a_B<a<1500a_B$. 
It shows that correlations diminish with increasing scattering
length. That is why the 
effective potential $\tilde V_{\rm p-h}(k)$
is completely determined by the bare two-body potential at large
$a$. In some sense that is equivalent to
the behavior of the charged
Bose gas \cite{bosegas} where the high density limit is the
weakly correlated fluid. 
%Yet, these systems behave differently at low
%densities where the long range of the Coulomb interaction leads to
%Wigner crystal. Here we have studied only short-ranged potentials with
%a Fourier transform, which do not have this gas-crystal phase
%transition and the system remains in the correlated, gaseous phase
%when $x\rightarrow 0$.

In comparison of different models we plot in
Fig. \ref{fig:effrangefunc} the s-wave effective range function $k\cot
(\delta_0(k))$ defined by the on-shell T-matrix
for the different models used.  At small $k<0.25$ it
is very close to -1 supporting the scattering length approximation.
At larger $k$ our model potentials give very similar downward bending.
At $k>0.6$ the first order Born approximation becomes very accurate
and we can conclude that the Fourier transform of the bare two-body
potential $\tilde V(k)$, determines completely the large $k$ behavior of the
effective range function.  
On the other hand, $\tilde V_{p-h}(k)$ for these momenta is also accurately
determined by $\tilde V(k)$. This is not surprising since at large
momenta the physics is governed by the short-distance two-body
problem.
This means that
% On the other hand, and as already pointed
% out, 
at large scattering lengths $\tilde V(k)$ could be determined by
the line shift measurements.  In Fig. \ref{fig:effrangefunc} we also
plot the hard sphere potential result and as shown it bends to the
wrong direction.
One can understand the link between $\tilde V_{p-h}(k)$
and the T-matrix by realizing that the large $k$ behaviour of both
quantities is determined by the 2-body physics at short distances

%seen it bends
%upwards whereas the line shift experiments require bending downwards.

Finally we give a simple parametrization by making a least square fit to the calculated effective range functions with a fourth degree polynomial
\be
k\cot (\delta_0(k)) =-1 +\half r_{\rm eff} k^2 + P k^4
\label{effra}
\ee
when $0<k<0.5$. The $k^4$ coefficient is the shape parameter $P$, which turns out to be important 
for reproducing well the bending of the line shift at large $a$.  With the PTGT-potential we find $P=-7.28$ and with the SW-potential
$P=-7.88$. The quality of the polynomial fits is also show in
Fig. \ref{fig:effrangefunc}.

%\begin{eqnarray}
%k \cot (\delta(k)) &=& -1 - 2.96 10^{-5}k^2 -7.28k^4
%\cr
%k \cot (\delta(k)) &=& -1 + 6.34 10^{-6}k^2-  7.88k^4
%\end{eqnarray}
%
In summary, we
%that within experimental resolution, the 
%line shift experiments by Papp et al. \cite{Papp2008} can be described
%using the static structure function and Feynman spectrum.  We 
have assumed that the effective two-body potential scales with the
scattering length and has a Fourier transform. Within these
assumptions many-body correlations diminish with increasing scattering length and 
line shift experiments could be used to  measure directly the Fourier transform of the
bare two-body effective potential. 
The downward bending of the line shift
clearly shows that 
%in alkali metal Bose gases the gas parameter $x$
%has now reached a level which goes beyond the zero range or hard
%core potentials, and more realistic models of the 
effective interactions with finite strength and range
are needed. By slightly increasing the gas parameter and momentum transfer one could study
experimentally the formation of a possible  density wave instability
%unavoidable if the Fourier transform of the bare, two-body 
%effective potential oscillates. 
and by increasing the
experimental resolution one could separate the elementary excitation
mode from the continuum of multiphonon contributions as seen in
$^4$He and the charged Bose gas. 
%These multiphonon contributions are
%responsible of the missing width in Papp et
%al. experiment\cite{Papp2008} as will be discussed in a forthcoming 
%work.

We thank V. Apaja, G. Astrakharchik, J. Boronat
 and A. Polls for discussions. One of us (R. S.) thanks the Finnish Cultural Foundation and Vaisala fund for financial support.
This work has been partially supported by Grant No. FIS2008-0443.

\bibliographystyle{apsrev}
% \bibliography{./papers_Bragg}

\end{document}